# Telecommunication wavelength confined Tamm plasmon structures containing InAs/GaAs quantum dot emitters at room temperature

Matthew Parker[1], Edmund Harbord[1]*, Lifeng Chen[1], Edmund Clarke[2], Kenneth Kennedy[2], John Rarity[1], Ruth Oulton[1]

[1]Quantum Engineering Technology Laboratories, H. H. Wills Physics Laboratory and Department of Electrical and Electronic Engineering, University of Bristol, Bristol BS8 1FD, United Kingdom
[2]EPSRC National Epitaxy Facility, Centre for Nanoscience and Technology, University of Sheffield, North Campus, Broad Lane, Sheffield S3 7HQ, United Kingdom
*edmund.harbord@bristol.ac.uk

**We experimentally demonstrate gold microdisc structures that produce confined Tamm plasmons (CTPs) – interface modes between a metal layer and a distributed Bragg reflector –resonant around 1.3 µm. Quantum dots grown within the structures show an order of magnitude increase in the photoluminescence emitted at room temperature. Varying the disc diameter, we show spectral tuning of the resonance and measure the dispersion relation as evidence of mode confinement. The simplicity of fabrication and tuneability of these structures make CTPs an ideal platform for making scalable telecom devices based on quantum dots.**

Epitaxially grown quantum dot (QD) heterostructures are a highly promising solid-state emitter for several devices. The zero-dimensional confinement of the charge carriers in QDs result in atom-like quantization of the energy levels for use as single-photon sources, which remain a necessary component for quantum information applications such as secure quantum communication [1] and quantum simulation [2]. They are bright, scalable, and have high internal efficiencies [3]. They are also highly tuneable: control of the growth conditions can modify the size and composition of the QDs so that they emit at the telecom wavelengths of 1.3 µm (O-band) or 1.55 µm (C-band) used for almost all optical communications [4-6]. The QDs emit photons in all directions within the GaAs host matrix, which results in most photons undergoing total internal reflection, and low external quantum efficiency. To make bright sources of single photons, it is necessary to engineer the photonic environment of the QD such that it emits into a single, highly collectable mode. Examples of such photonic structures include micropillar cavities [7] or the nearfield of a surface plasmon [8].

Optical Tamm states are confined electromagnetic modes that occur at the interface of photonic crystals: they are optical equivalents to the electronic states at the surface of crystal lattices originally described by Tamm [9]. When the photonic crystal is terminated by metal layer, the resulting Tamm optical state is known as a Tamm plasmon (TP) [10,11]. 1D Tamm plasmons can be formed between a Distributed Bragg Reflector (DBR) and a thin (nanometer-scale) metal layer. Decaying solutions are produced within the stopband of the DBR (as these produce imaginary values of the Bloch wavevector) and within the metal due to its negative permittivity. TPs therefore produce a dispersion relation within the light line $\omega = c\kappa$ and unlike surface plasmons require no additional mechanism to couple with freely propagating light. They are excited by normal incidence light and in both TE and TM polarizations [12]. However, their most attractive feature is the option to control the mode resonance and emission by altering the metal layer [13-16]. Depositing a metal layer with micron-scale dimensions causes the TPs to be confined laterally to make confined Tamm Plasmons (CTPs). The benefits of fully 3D confinement of the mode can hence be realized using a relatively simple and scalable metal deposition techniques, and which avoids the need for destructive etching. These properties have led to growing interest in TPs being used in optical devices such as optical switches and transistors [17] or detectors [18].

QDs positioned within the confined fields of a TP or CTP have increased brightness and highly directional emission [13]. This has been used to decrease the lasing threshold for QD lasers [19] and improve collection efficiency for single-photon sources [20,21]. Asymmetric structures have been used to fabricate nanolasers [14] and for polarization resolved photodetection [22]. In this paper we present CTP structures consisting of a gold microdisc resonant at 1.3 µm and show evidence of QDs coupling into CTPs at telecoms wavelength. Photoluminescence (PL) collection through the top of the structure increases by an order of magnitude compared to when the CTP is absent and shows evidence of mode confinement when disc diameter is reduced, which also presents a path for spectral tuning of the mode. These devices have potential uses for single-photon sources, lasers and photodetectors for telecom applications.

CTP structures were previously calculated to produce resonant modes at the desired wavelength using the transfer matrix method (TMM) for planar (i.e., non-confined) TPs and the finite-difference time-domain (FDTD) method for microdisc structures [23]. A TP resonant at 1.3 µm is predicted for a structure consisting of a 17.5x λ/4 GaAs/AlAs pair DBR, a 75 nm GaAs layer, called the spacer layer, and a 25 nm gold layer. To meet the phase matching conditions of a Tamm mode the DBR must have a positive reflection phase, meaning this spacer must be the higher of the two refractive index materials [11,24]. The reflectivity spectrum calculated by the TMM is shown in figure 1(a) without (blue line) and with (red line) the metal layer; for the DBR only the photonic stopband appears as the window of near-unity reflectivity, while the addition of the metal layer creates a TP resonant at $\lambda_{TP}$ = 1304 nm, seen as the dip in reflectivity where light couples into this mode. Calculated electric field intensities at this wavelength are shown in profile through the structure in figure 1(b). The TP is seen as an enhancement of the field compared to that of the bare DBR occurring within the spacer layer.

Lateral confinement is added by limiting the dimensions of the metal layer. This is shown schematically in figure 1(c). Increasing the confinement results in a blueshift of the Tamm resonances. When confined the TP's dispersion, which in the planar case is a distinctive continuous parabola, is discretized with increased spectral separation between the different Tamm resonances [13,25]. FDTD simulations of the structures previously described with the gold layer replaced with gold microdisc of diameter $d$, also produce these effects [23]. A vertical cross-section for a $d$ = 2.6 µm CTP is shown in figure 1(d) which shows the electric field is confined by the lateral extent of the disc (indicated by the white lines) and is strongest under the disc center.

Samples are grown using molecular beam epitaxy (MBE) on a GaAs substrate, consisting of a 17.5x GaAs/AlAs pair DBR with thickness of 94.5/110.8 nm and a 75 nm GaAs spacer. Gold microdiscs of various diameters between 0.8 µm to 13 µm are deposited using a photolithographic resist and lift-off technique. A 20 µm disc is used to emulate the planar TPs. These are deposited for a range of metal thicknesses by DC sputtering. The thickness of the metal was determined by ellipsometry measurements of metal layers deposited on planar silicon reference samples at the same time and co-located with the Tamm sample in the sputtering chamber and was confirmed by measurements of some Tamm disc samples by atomic force microscopy. An optical microscopy image of the discs is shown as an inset in figure 4(a).

The samples are characterized using Fourier image spectroscopy (FIS), as previously described in Ref. 26. The image at the back focal plane of the objective lens contains the spatial Fourier transform of the sample's optical response, since every ray with the same spatial frequency is focussed by the lens to the same position at this plane. Spectra are taken with a multimode (200 µm core) fiber leading to an Ocean Optics NIRQuest512 spectrometer. The fiber is mounted to a motorized translation stage that raster scans through several points in the Fourier image. This enables angle-resolved measurements of the reflectance and PL for different wavelengths, allowing the dispersion of the TPs to be obtained. We can use either an unpolarized white light source for reflectivity measurements, or 635 nm laser diode for photoluminescence measurements. These are focussed onto the sample by an objective lens. The image at the back focal plane of the objective is then relayed through a standard 4$f$ system to the detection plane containing the mounted fiber and spectrometer. The angular response of the system is calibrated by reflectivity measurements of a 1000 nm period linear grating, so every spatial position of the fiber in the back focal plane can be given a corresponding angle. Excitation and collection both occur through the metal side of the structure.

The normal incidence reflectivity spectrum for a large microdisc ($m$ = 23.1 nm, $d$ = 20 µm) is shown in figure 2(a) (orange line). This is sufficiently large with respect to the wavelength that we can model it as a 1D cavity. A TP mode can be observed as the dip in reflectivity occurring at $\lambda_{TP}$ = 1302 nm. Reflectivity of the bare DBR (green line) and from the TMM simulation (red and blue lines) are shown for comparison. The measured results have close agreement with the TMM for both the position of the DBR stopband and the TP respectively. The reflectivity of the measured DBR stopband is greater than unity as it is more reflective than the gold mirror reference to which it is normalized. As the layer thickness is decreased the wavelength of the reflectivity dip shifts to lower energies (figure 2(b); black squares). This also shows strong agreement with that predicted for planar TPs. Further increase of the metal thickness has a limited effect on the reflection phase imparted (and hence position of the resonance condition, for a given DBR) resulting in the asymptotic behaviour at thicker layers. However, because these are excited through the metal layer this decreases the amount of light, through reflection and absorption, from exciting the TP. This decreases the magnitude of the reflectivity dip, which could not be observed for thicknesses in excess of 40 nm.

For PL measurements inclusions of InAs QDs are grown within a 5 nm layer of $In_{0.18}Ga_{0.82}As$, which acts as a quantum well. Growing within the well maintains the size and composition of the QDs during capping and reduces the strain on the dots, resulting in a redshift of their emission that allows telecom wavelengths to be reached [4, 27]. The PL sample contains four layers of InAs/InGaAs dot-in-wells; first, at 5 nm from the AlAs-spacer interface, predicted to coincide with the maximum field intensity of the TP, and at the next three subsequent antinodes, each a single DBR period in depth from the previous layer (the sample used in the reflectivity measurements, which contained only the first dot-in-well layer, did produce PL but was not bright enough for low-noise measurement). The QDs are undoped and have a density of 5.7 x $10^{10}$ cm$^{-2}$ per layer. A PL spectrum taken before deposition of the metal layer is shown in figure 3(a). This has a broad emission with a FWHM between 1290 nm and 1330 nm and the emission peak occurring at 1312 nm, as measured using an Accent RPM 2000 PL mapper.

Room temperature PL of the samples are taken using a continuous wave 635 nm laser source (above the GaAs bandedge). The PL emission of the dots act as a broad light source centered at 1310 nm, which is filtered by the enhancement of emission resonant with the TP. Figure 3(b) shows such a spectrum for a $m$ = 23.1 nm, $d$ = 20 µm microdisc (blue line) with the result off-disc, i.e. bare DBR, shown for comparison (black line). In both cases a broad emission at shorter wavelengths is observed, owing to excited state filling and cavity enhancement effects (this appears larger in the spectrum taken with the Fourier microscope owing to a higher degree of state filling associated with the smaller spot size of this set-up). However, when the metal layer is present a distinct enhancement is seen, corresponding closely to the QDs emission peak, but shifted and narrower. Angle-resolved reflectivity and PL is shown in figure 3(c). The dispersion for the $d$ = 20 µm disc, shown by the reflectivity minima, has a continuous and parabolic dispersion expected for planar TPs. The pattern of PL enhancement at higher angles follows the dispersion relation of the TP, demonstrating that emission from QDs coupling to this mode is the cause of the enhancement.

To separate the TP from this spectrum the sideband of the QDs is fitted to a skewed distribution and subtracted. The resonance spectral position and Q factor are obtained from a Gaussian fit to this remaining response. For the $m = 23.1$ nm disc, shown in figure 3(a), there is an increase in PL collected by x11 compared to the bare DBR. To test the PL dependence on pumping power, neutral density (ND) filters of increasing optical density are added to the excitation path. This showed that at our maximum laser power (0.80 mW) the enhancement does not cause saturation of the QDs. The dependence of Q-factor with metal thickness is shown in figure 3(d). However, though the Q-factor is shown to increase with thicker metal layers there is a decrease in the PL emission, as the reduction of emission collected through the top of the structure from absorption and reflection as the metal layer is increased offsets any increase in confinement of the TP. The strongest enhancement therefore occurs for the $m = 23.1$ nm CTP.

We report the effect of microdisc diameter on the PL for disks of metal thickness 23.1 nm. PL is next measured for microdiscs with varying diameter. The shift in peak position as a function of diameter is shown in figure 4(a). As the diameter is reduced, there is a blueshift in the Tamm resonances, showing the effect of confinement of the mode. There is also a general trend of decreasing Q-factor (figure 4(b); red line). This same trend is seen when FDTD simulations of fully confined TPs are performed (blue line), though there are also oscillations seen in the experimental results not previously predicted. The measured Q-factor for the $d = 20$ μm CTP is 20% lower than the expected value for the non-confined case. The Q-factor difference between experiment and FDTD decreases at smaller diameters. We attribute this initial lowering of the Q-factor to effects such as metal layer roughness which are not included in FDTD; at smaller diameters scattering and plasmon excitation occurring at the disc edge, which are included in FDTD, become the dominant loss mechanism, and hence there is a closer agreement between the two values.

Confinement is also demonstrated by changes in the dispersion relation seen in the angle-resolved PL of smaller diameter discs (figure 4(c)). For the $d = 13$ μm CTP the dispersion indicated by the PL enhancement is still continuous. In the $d = 2.6$ μm case a much flatter dispersion relation which is blueshifted from the continuous case heralds the onset of confinement. This value agrees well with the predicted shift for the fundamental (i.e. $LP_{01}$) CTP mode previously predicted using an effective index model [28] for a 2.6 μm disk.

In conclusion, we have demonstrated microdisc structures that produce confinement of Tamm plasmon modes resonant at telecoms wavelength. These can be fabricated with relatively simple metal fabrication techniques. Shift in CTP resonance and Q-factor with metal layer thickness and disc diameter agrees well with TMM and FDTD simulations, the former allowing spectral tuning of the device. Fourier image spectroscopy of the PL shows QD coupling to the CTP as an enhancement in collected emission. For our optimum structure this produces an enhancement of x11. The scalability, tunability and robustness of the CTPs make them an ideal platform for the development of novel detectors, single-photon sources and optical switches at telecoms wavelength.

This work was partially supported by ESPRC grants EP/M024156/1, EP/N003381/1, andEP/M024458/1 .

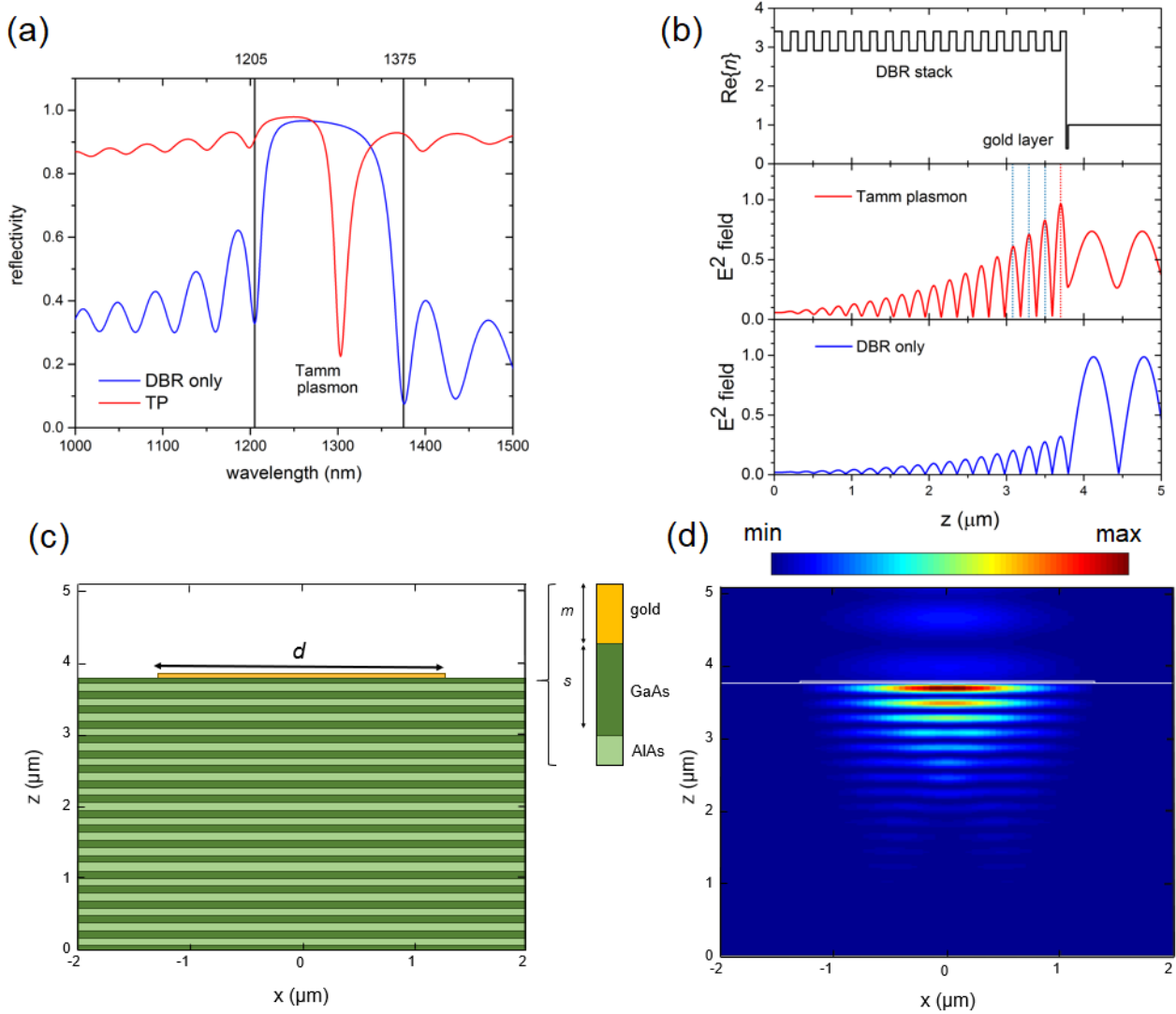

**FIG 1. (a)** Power reflectivity coefficients at normal incidence calculated using the transfer matrix method (TMM) for the planar (i.e. nonconfined) structure consisting of a 17.5x pair GaAs/AlAs DBR, 75 nm GaAs spacer layer and with (red) or without (blue) a 25 nm gold layer. The photonic stopband of the bare DBR is shown as a window of high reflectivity between 1205 nm and 1375 nm. With the metal layer, the TP is seen as a reflectivity dip occurring at $\lambda_{TP}$ = 1300 nm. **(b)** Cross-section of the refractive index and TMM calculated electric field at 1300 nm with (red) and without (blue) the metal layer. The dashed lines show the position of the first four antinodes of the Tamm mode where the InAs/AlAs QD layers are deposited. **(c)** Diagram of a CTP structure consisting of a diameter $d$ gold microdisc. **(d)** Profile of an FDTD simulation showing the absolute electric field intensity at 1300 nm for the structure shown in (c) (m = 25 nm, d = 2.6 μm) excited from above the microdisc (marked in the figure by white lines).

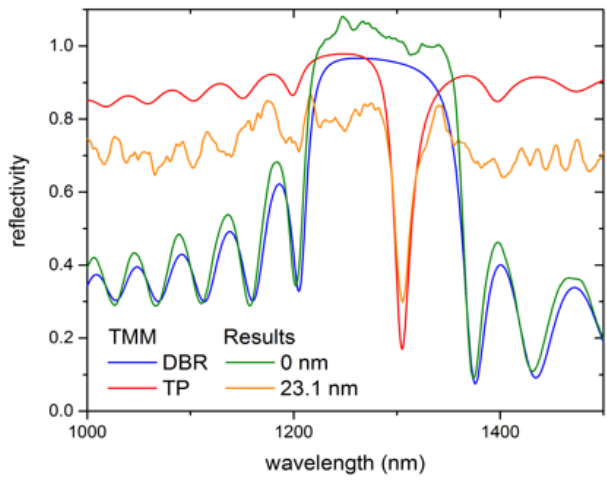 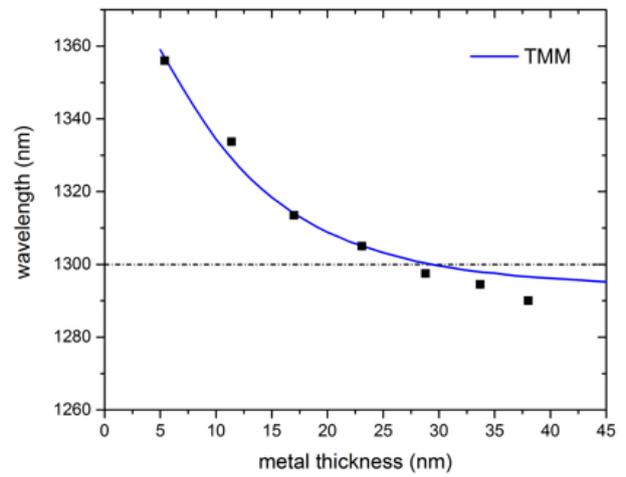

*FIG 2. (a) Normal incidence reflectivity spectrum measured for the bare DBR (green line) and d = 20 μm CTP structure (orange line). The TMM calculation for the DBR and m = 23.1 nm is shown for comparison by the blue and red lines. (b) Position of the reflectivity dip as a function of metal thickness (black squares) for the d = 20 μm discs, with the TMM predicted resonance shift (for a planar TP) shown by the blue line.*

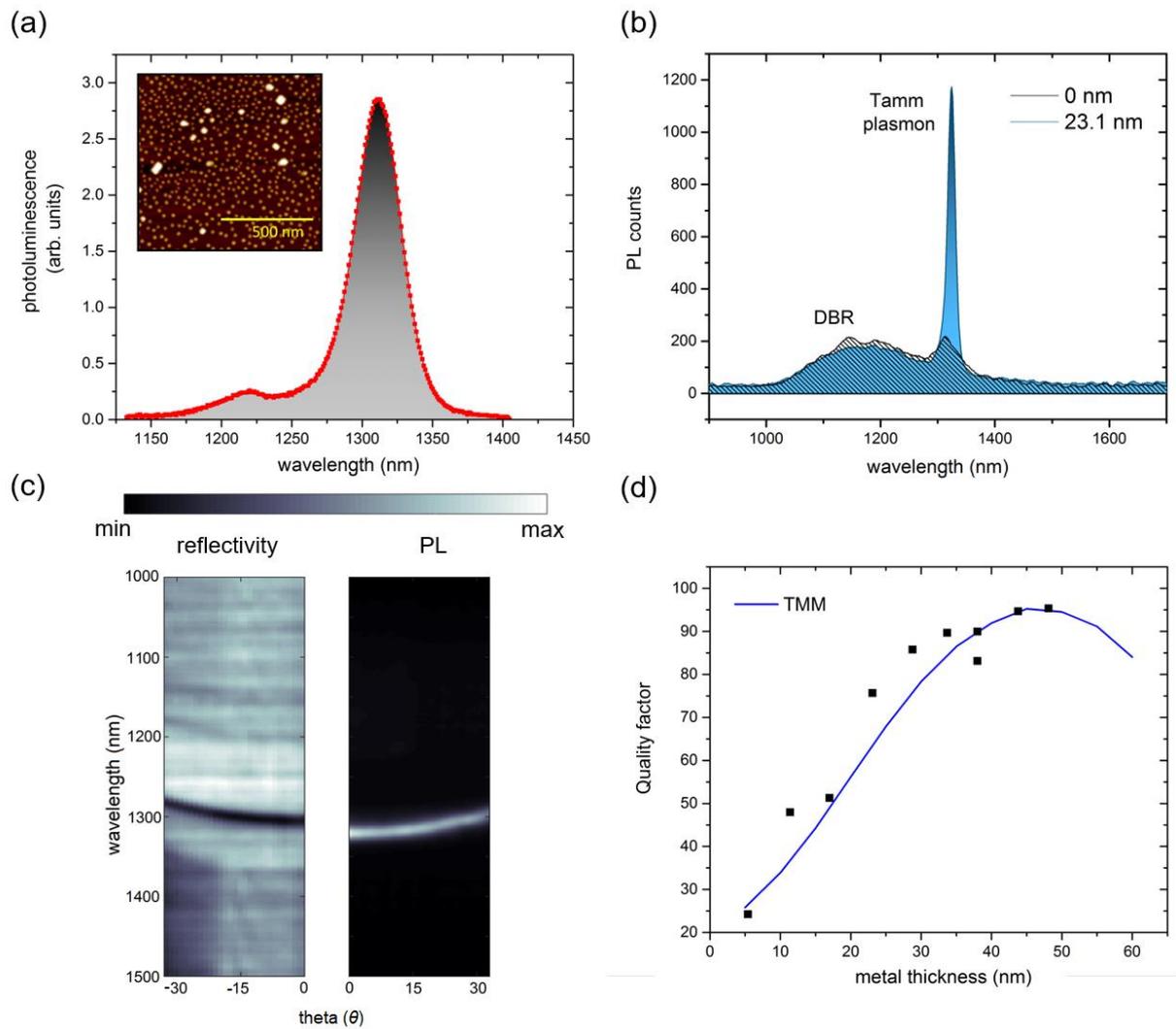

*FIG 3. (a) Photoluminescence spectrum of the sample before deposition of the metal layer (bare DBR only) showing the room temperature emission from the InAs/GaAs QD layers. The emission peak occurs at 1312 nm with a FWHM of 40 nm. The additional peak at 1215 nm is attributed to enhancement from the first Bragg mode of the DBR.* **Inset:** *AFM measurement of uncapped QDs grown using the method described. (b) PL spectra of the* m = 23.1 nm, d = 20 μm *sample taken in the Fourier microscope at normal incidence. The QDs are excited non-resonantly at 635 nm. The black and blue lines show the PL collection when focused off and onto the microdisc respectively. (c) Angle-resolved PL measurement taken with the FIS using a 40x (NA = 0.75) objective. Angle-resolved reflectivity is shown for comparison. (d) Q-factors of the normalized PL peaks as a function of metal thickness (black squares) and values calculated using the TMM (blue line).*

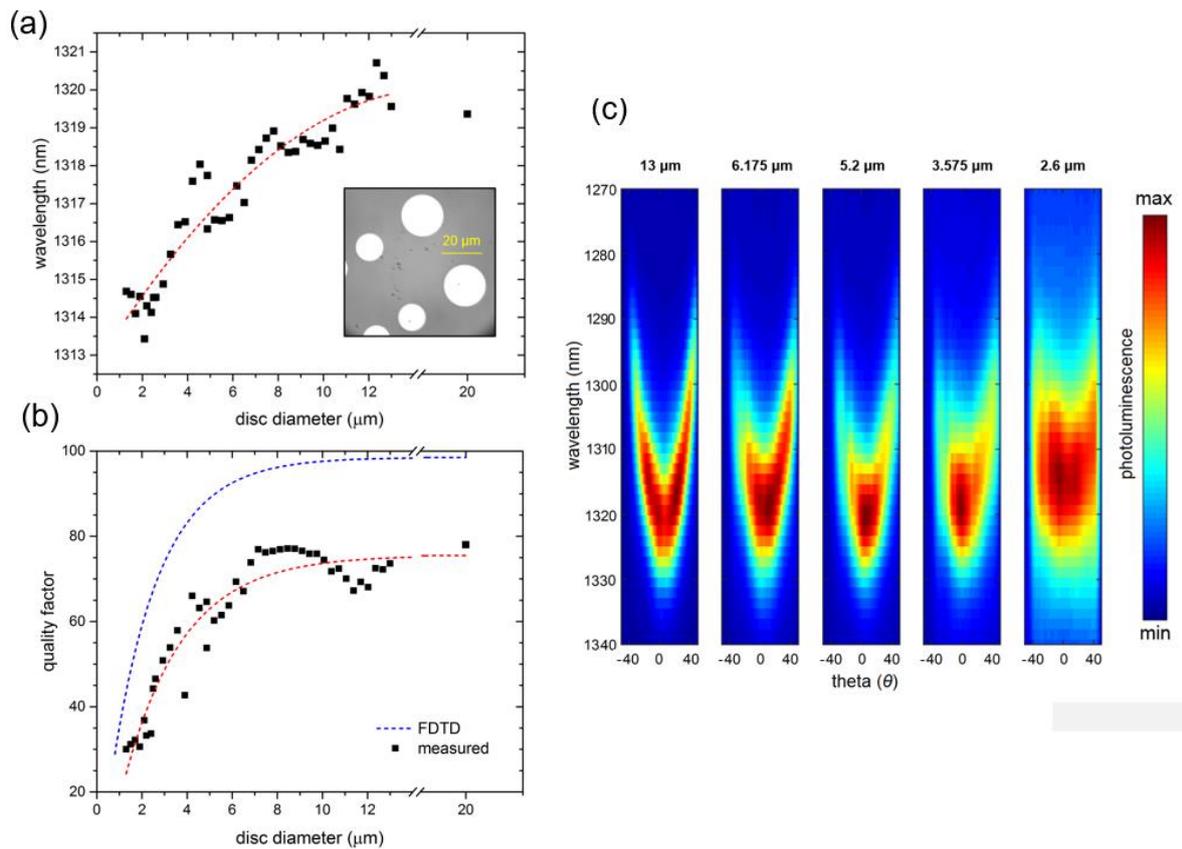

*FIG 4. (a) Resonance and (b) Q-factors as a function of disc diameters, taken from the normalized PL spectra for CTP structures (m =23 nm). The red dashed line in (a) is a guide to the eye. The red and blue dashed lines in (b) are exponential functions fitted to the measured and FDTD results respectively. **Inset:** Real image of the microdiscs taken with a 40x objective. (c) Angle-resolved photoluminescence (PL) spectra for different disc diameters taken with a 100x (NA = 0.8) objective. As the diameter is reduced there is a transition from a continuous and parabolic dispersion, occurring in planar TPs, to a discrete mode centered at the CTP fundamental resonance.*